\documentstyle[sprocl]{article}

\newtheorem{theorem}{Theorem}[section]

\newtheorem{lemma}[theorem]{Lemma}

\newcommand{\IR}{{\hbox{\rm \  I\hskip-1.4pt R}}}
\newcommand{\NN}{\rm \  N \kern-0.85em N\,}

\begin{document}
\title{A LOCAL MATHEMATICAL MODEL FOR EPR-EXPERIMENTS}
\author{W. Philipp}
\address{Beckman Institute\\Department of Statistics and Department of Mathematics\\
University of Illinois, Urbana, IL 61801\\E-mail:
wphilipp@uiuc.edu}

\author{K. Hess}
\address{Beckman Institute\\Department of Electrical Engineering and Department of
Physics\\University of Illinois, Urbana, IL 61801\\E-mail:
k-hess@uiuc.edu}
\maketitle

\abstracts{In this paper we give a detailed and simplified version
of our original mathematical model published first in the
Proceedings of the National Academy of Science.  We hope that this
will clarify some misinterpretations of our original paper.}
\medskip \medskip

\section{Introduction}\label{sec1}
In \cite{hp,hpp1,hpp2}, and \cite{hpp3} we presented a local
mathematical model for EPR-type experiments. Our model is in
agreement with the results predicted by Quantum Mechanics which in
turn were confirmed by large scale experiments, first by Aspect,
Roger and Dalibard~\cite{eprex} and later by several other teams.
Due to space limitations the presentation of our model was rather
terse in places. The purpose of this paper is to present our model
in much greater detail and at the same time mathematically
simplified because concerns and questions have been raised about
non-locality and parameter dependence in a few recent
publications~\cite{myr,gwzz,gwzz1,gwzzpnas,apy}. Although we have
answered these concerns at various
occasions~\cite{hpm2,gref,apyref}, a comprehensive detailed
exposition might be a better way to address these concerns.

At first we give a brief summary of our model. In EPR type
experiments two particles having their spin in a singlet state are
emitted from a source and are sent to spin analyzers at two
spatially separated stations $S_1$ and $S_2$. We assume with Bell
that the particles emitted from the source are permitted to carry
information in form of arbitrary hidden parameter random variables
$\Lambda$ that can assume values in some abstract space.

In the original experiment by Aspect \cite{eprex}, pairs of
photons were emitted from the source once every few microseconds
over a hour period governed by a random process. We model this
process mathematically in the following way. Imagine the time axis
wrapped around a circle of circumference that corresponds to a
time interval related to a simple measurement and normalized to 1.
We suppose that for a fixed $N$ each interval $[(m-1)/N, m/N]$, $m
= 1, 2, \ldots, N$  of arc length $1/N$ on the circle gets about
its proper share of time measurement points over the measurement
period. This assumption, in turn, induces a random variable $R$,
that we call the labelling variable, which assumes the values $m =
1, 2, \ldots, N$ with equal probability, i.e.,
\begin{eqnarray}
P (R = m) = 1/N & \quad & m=1,2, \ldots, N.
\label{eq1}
\end{eqnarray}
In Section~\ref{sec4} below, we present better motivation for the
generation of the labels $m$ and the random variable $R$ by means
of the Poisson process, commonly used to model spontaneous
emissions.

In our papers~\cite{hp,hpp1,hpp2,hpp3}, using notation standard in
Bell type proofs, the source parameter now denoted by $\Lambda$
was not assigned a separate letter. The letter $\lambda$, which in
the standard notation generically symbolizes randomness, was used
instead.  As a consequence the label $m$ being a function of the
random emission times was misinterpreted by various authors as a
function of $\lambda$. In the present paper we have therefore
decided to use the standard probability notation instead: Random
variables are given separate names, they will be denoted by
capital letters, they are measurable functions of $\omega$ or
$\lambda$, attached to some experiments, and there will be a clear
distinction between the random variables, the values they can
assume, and the set of measured data.

After a pair of particles has been emitted from the source, the
time of emission and thus of measurement are known, and so is the
interval $[(m-1)/N, m/N]$ on the unit circle into which the time
of measurement falls. This determines the label $m$. While the
pair of particles travels to their designated analyzer stations
the experimenters (or a random number generator) can exercise
their free will and choose in their respective stations the
directional settings, say $\textbf{a}$ in $S_1$ and $\textbf{b}$
in $S_2$. Our model calls for hidden parameter random variables
\cite{foot1} $\Lambda^*_{{\bf a}t}$ in $S_1$ and
$\Lambda^{**}_{{\bf b}t}$ in $S_2$ which depend on the respective
settings and on the time of measurement.  The time of measurement
is known (either the same or connected by a linear relation) in
both stations. Consequently, the label $m$ is known at both
stations. This provides for the time correlation we elaborated in
papers~\cite{hp,hpp1,hpp2} and \cite{hpp3}.

Thus we have four random variables in operation, $R$, $\Lambda$,
$\Lambda^*_{{\bf a}t}$, and $\Lambda^{**}_{{\bf b}t}$. The joint
density of all these variables $\rho_{\bf ab}$, is
permitted~\cite{foot2} to depend on the settings $\textbf{a}$ and
$\textbf{b}$, and is given in Eq.~(\ref{eq29}). As a consequence
of our construction we obtain certain properties of stochastic
dependence relations which we state as a preview. In the general
case where the distribution of $\Lambda$ may depend on time we
have the following stochastic dependence relations between these
four random variables signifying their time correlations.
\begin{description}
    \item{(i)} The random variables $\Lambda_{{\bf a}t}^*$ and
    $\Lambda_{{\bf b}t}^{**}$ are stochastically independent.
    \item{(ii)} Given the random variable $R$ the pair
    $(\Lambda_{{\bf a}t}^*, \Lambda_{{\bf b}t}^{**})$ is conditionally
    independent of $\Lambda = \Lambda_t$.
\end{description}
As to the probability distributions our construction yields the
following properties.
\begin{description}
    \item{(iii)} The probability distribution of $\Lambda_t$ can
    be chosen arbitrary.
    \item{(iv)} The probability distributions of $\Lambda_{{\bf
    a}t}^*$ and $\Lambda_{{\bf b}t}^{**}$ do not depend on ${\bf a}$,
    ${\bf b}$, nor $t$.
\end{description}
In the special case where the distribution of $\Lambda$ does not
depend on time we have in addition to the above properties
\begin{description}
    \item{(ii)$^*$} The random variables $\Lambda$, $\Lambda_{{\bf
    a}t}^*$, and $\Lambda_{{\bf b}t}^{**}$ are stochastically
    independent.
    \item{(vi)$^*$} The random variables $R$ and $\Lambda$ are
    stochastically independent.
\end{description}

The random variables $A_{\bf a} = \pm 1$ and $B_{\bf b} = \pm 1$
symbolize the possible spin values and are functions {\it only} of
${\bf a}$, $\Lambda$, $R$, and $\Lambda_{{\bf a}t}^*$, and of
${\bf b}$, $\Lambda$, $R$, and $\Lambda_{{\bf b}t}^{**}$,
respectively, thus obeying Einstein locality; that is
\begin{eqnarray*}
A_{\bf a} & = & A_{\bf a} (\Lambda_{{\bf a}t}^*, \Lambda; R)
\end{eqnarray*}
and
\begin{eqnarray*}
B_{\bf b} & = & B_{\bf b} (\Lambda_{{\bf b}t}^*, \Lambda; R).
\end{eqnarray*}

Moreover, we have
\begin{eqnarray}
E \{A_{\bf a} B_{\bf b}\} & = & - {\bf a} \cdot {\bf b} = - \cos
<) ({\bf a}, {\bf b})
 \label{eq2}
\end{eqnarray}
and thus we have with probability 1
\begin{eqnarray}
B_{\bf a} & = & - A_{\bf a}.
\label{eq3}
\end{eqnarray}
In addition, we have with probability 1
\begin{eqnarray}
E \{A_{\bf a} \mid \Lambda, \Lambda_{{\bf a}t}^* \} & = & E
\{B_{\bf b} \mid \Lambda, \Lambda_{{\bf b}t}^{**} \} = 0.
 \label{eq4}
\end{eqnarray}
Note that the integration is, in essence, only performed with
respect to $R$.  If, in addition we integrate Eq.~(\ref{eq4}) with
respect to $\Lambda_{{\bf a}t}^*$ and $\Lambda_{{\bf b}t}^{**}$ we
obtain with probability 1
\begin{eqnarray}
E \{ A_{\bf a} \mid \Lambda \} & = & E \{B_{\bf b} \mid \Lambda \}
= 0. \label{eq5}
\end{eqnarray}
Since $A_{\bf a} = \pm 1$ and $B_{\bf b} = \pm 1$, Eq.~(\ref{eq5})
implies parameter independence.  However, our variables $A_{\bf
a}$ and $B_{\bf b}$ depend on time through their functional
dependence on $R$, a feature that is not considered in the
original Bell definition of $A_{\bf a}$ and $B_{\bf b}$.

Thus our model is distinguished from standard Bell-type models by
the introduction of the time related labelling variable $R$. If in
Eq.~(\ref{eq5}) we also condition on $R$, besides conditioning on
$\Lambda$, these equations no longer will hold. Moreover, we note
that in the general case we have:
\begin{description}
    \item{(v)} Given the random variable $R$ the random variables
    $\Lambda_{{\bf a}t}^*$ and $\Lambda_{{\bf b}t}^{**}$ are
    stochastically dependent.
    \item{(vi)} The random variables $R$ and $\Lambda_t$ are
    stochastically dependent.
    \item{(vii)} The conditional probability distributions of
    $\Lambda_{{\bf a}t}^*$ given $R$ and of $\Lambda_{{\bf b}t}^*$
    given $R$ depend on both settings ${\bf a}$ and ${\bf b}$.
\end{description}

On a more basic level probabilities, conditional probabilities or
even conditional expectations, such as the one in Eq.~(\ref{eq5}),
can be interpreted as long term averages of outcomes of certain
experiments.  These long term averages can be thought of being
taken over certain points on the time axis. We separate this
averaging process into two parts by introducing the random
variable $R$.  We first average over the concatenated time
intervals associated with a fixed label $m = 1, 2, \ldots, N$.
Subsequently, we average the first averages over the values $m$
that $R$ can assume to obtain the overall averages.

As a consequence we do not view the conditional stochastic
dependence in (v), nor the dependence on the settings of the
conditional probability distributions in (vii) as a violation of
Einstein locality.  These dependencies only signify the time
correlations between the events in stations $S_1$ and $S_2$. To
express this in physical terms we point the reader to the
following facts. The label $m$ represents a concatenation of short
time segments and not a given time. $m$ therefore does not relate
to or permit any instantaneous signalling. It can not be
influenced by the experimenter in any significant way since it
depends on the random spontaneous emission times and the largely
arbitrary way of concatenating these short time intervals in a
specific interval $[(m-1)/N, m/N]$. Therefore $m$ is not an
element of reality as opposed to, for example, the source
parameter $\Lambda$.

Any argument for instantaneous action at a distance involving
probabilities conditional on $\{R=m\}$ must therefore be
counterfactual. As an important example, one could argue that
instead of a setting pair ${\bf a}, {\bf b}$ the experimenters
might have chosen ${\bf a}, {\bf c}$. Then, since the joint
probability conditional on $\{R=m\}$ depends on both settings, the
marginal distribution of $\Lambda_{{\bf a} t}^*$ for setting ${\bf
a}$ conditional on $\{R=m\}$ may be different.  How can this be
without instantaneous action at a distance? The answer is that if
${\bf c}$ would have been chosen, then over a whole sequence of
measuring times all the settings would be different. In order to
have setting ${\bf b}$ with equal probability to setting ${\bf
c}$, the experimenters would have had to decide to choose ${\bf
c}$ instead of ${\bf b}$ at other occasions. In other words, the
whole history of settings would have to be different. Because all
the involved parameters, as well as the possible outcomes for the
spin pair values may depend on the history, the probability
distribution of $\Lambda_{{\bf a}t}^*$ conditional on $\{R=m\}$
may depend on the history and can therefore be different for the
setting pairs ${\bf a}, {\bf b}$ and ${\bf a}, {\bf c}$. The EPR
argument postulates a physical reality of the source parameter
$\Lambda$; in our papers we postulate also physical reality for
the station parameters $\Lambda_{{\bf a}t}^*$ and $\Lambda_{{\bf
b}t}^{**}$. However, we do not attach a physical reality in the
same sense to the labelling random variable $R$. Fulfillment and
violation of Einstein locality with respect to random variables
such as the labelling variable $R$ becomes a highly complex
problem \cite{foot3}. Let us note, in passing, that the exclusion
of setting dependence conditional to any concatenation of time
segments such as represented by $m$ will automatically also
exclude the result of the actual experiments which can be regarded
as performed by concatenating the results obtained in certain time
segments.

We would like to emphasize that the joint probability measure
given by Eq.~(\ref{eq29}) below is not canonical, i.e., not
unique. This makes the model highly flexible to accommodate other
possible set-ups of experiments. In fact, we hope to show with our
work that the choice of the particular form of variables is
mathematically highly flexible and can go far beyond simple ideas
of elements of physical reality.  We do not claim that the
particular model actually exists in nature. All we want to show is
that Bell type proofs actually can not do justice to the
complexities involved in EPR experiments and therefore can not be
used to draw conclusions about nonlocal effects as epitomized by
instantaneous action at a distance.

For clarity of presentation we develop our construction in several
steps.  The first one, almost identical with the presentation
in~\cite{hp,hpp2}, will be given in Section~\ref{sec2}. In
Section~\ref{sec3} we define the probability distribution of the
time and setting dependent station parameters $\Lambda^*_{{\bf
a}t}$ and $\Lambda^{**}_{{\bf b}t}$.  The measure we construct in
these sections is not quite a probability measure (see
Eq.~(\ref{eq10}), below).  However, it is a routine exercise to
derive from it a probability measure by applying some basic facts
from the theory of weak convergence of probability measures.  We
may present the details in a paper to be submitted to a
mathematics journal.

We hope that this introduction provides enough of a guiding line
through the mathematical intricacies that will follow.

\section{The First Step in Establishing the Model} \label{sec2}
Before we start with the mathematics, let us recall that a pair of
particles has been emitted from the source. The emission time and
thus the measuring time is known. As a consequence the value $m$
of the labelling variable $R$ is determined.  The experimenters
have subsequently chosen their vectors $\textbf{a}$ in $S_1$ and
$\textbf{b}$ in $S_2$, respectively. In effect, we assume that the
measuring time, considered as a random variable and the labelling
variable $R$ are independent of the choice of vectors $\textbf{a}$
and $\textbf{b}$.

Let $\textbf{a} = (a_1, a_2, a_3)$ and $\textbf{b} = (b_1, b_2,
b_3)$ be unit vectors.  Our goal is to show that under our
generalized conditions, it is possible to obtain the quantum
result, the scalar product $-\textbf{a} \cdot \textbf{b}$ for the
spin pair expectation value $E \{ A_{\bf a} B_{\bf b} \}$. Here we
formulate a theorem which provides the first stepping stone for
this procedure.

We define functions $A_{\bf a}$ and $B_{\bf b}$ and choose the
underlying measure space $(R^2, {\cal{R}}^2)$, i.e., the Euclidean
plane $\{(u,v), - \infty < u, v < \infty \}$ with Borel
measurability, symbolized by ${\cal{R}}^2$.  We set
\begin{eqnarray}%
A_{\bf a} (u) & = & \left\{ \begin{array}{lll} {\mbox{sign}} (a_k)
& \mbox{ if } - k \leq u < - k + 1 & k = 1, 2, 3 \\ -1 & \mbox{ if
} j \leq u < j + {1 \over 2} & j = 0, 1, \ldots\\ +1 & \mbox{ if }
j + {1 \over 2} \leq u < j + 1 & j = 0, 1, \ldots \\ +1 & \mbox{
elsewhere}. &
\end{array} \right.
\label{eq8}
\end{eqnarray}
Thus $A$ depends here on $\textbf{a}$ and $u$ only.  We will
return below to the complete list of dependencies which only here
would complicate the notation and not add to the present purpose.
Here and throughout, we set ${\mbox{sign}} (0) = 1$. Similarly, we
define
\begin{eqnarray}
B_{\bf b} (v) & = & \left\{ \begin{array}{lll}
- {\mbox{sign}} (b_k) & \mbox{ if } - k \leq v < - k + 1 & k = 1, 2, 3 \\
+1 & \mbox{ if } j \leq v < j + {1 \over 2} & j = 0, 1, \ldots \\
-1 & \mbox{ if } j + {1 \over 2} \leq v < j + 1 & j = 0, 1, \ldots
\\ -1 & \mbox{ elsewhere}. & \end{array} \right.
\label{eq9}
\end{eqnarray}
As in the case for $A$ above $B$ depends for the moment on ${\bf
b}$ and $v$ only.  We now formulate the first step as a theorem.

\begin{theorem}\label{thm2.1}
Let $n \geq 4$ be an integer.  Then there exists a finite measure
$\mu = \mu_{\bf ab}^{(n)}$ with the following properties: $\mu$
depends only on $n$, ${\bf a}$ and ${\bf b}$, has compact support
$\Omega$, satisfies
\begin{eqnarray}
1 & \leq & \mu (R^2) \ \ < 1 + 1/n^2
\label{eq10}
\end{eqnarray}
and has a density $\rho = \rho_{\bf ab}^{(n)}$ with respect to
Lebesgue measure. Further
\begin{eqnarray}
\int_{\Omega} A_{\bf a} (u) B_{\bf b} (v) \rho_{\bf ab}^{(n)} (u,
v) du dv & = & - {\bf a} \cdot {\bf b}
\label{eq11}
\end{eqnarray}
and for each vector ${\bf a}$ the following equation holds for all
$x$:
\begin{eqnarray}
B_{\bf a} (x) & = & - A_{\bf a} (x).
\label{eq12}
\end{eqnarray}
\end{theorem}

The proof of the theorem requires the following fact which follows
from a basic theorem on $B$-splines~\cite{sch81}.  We state the
fact here in form of a lemma.

\begin{lemma}  Let $n \geq 4$ be an integer.  Then
there exist real-valued functions $N_i (x)$, $\psi_i (y)$ with $1
\leq i \leq n$ depending only on real variables $x$ and $y$,
respectively, such that
\begin{eqnarray}
0 \leq N_i (x) \leq 1, &&
0 \leq \psi_i (y) \leq 2 \mbox{ for } 0 \leq x,y \leq 1%
\label{eq13}
\end{eqnarray}
and
\begin{eqnarray}
0 & \leq & \sum_{i=1}^n \psi_i (y) N_i (x) - (y-x)^2 \leq {1 \over
4} n^{-2} \mbox{ for } 0 \leq x, y \leq 1.%
\label{eq14}
\end{eqnarray}
\end{lemma}
The proof of this lemma is given in Appendix~1.  We now proceed to
prove Theorem~\ref{thm2.1}.\bigskip

\paragraph{Proof of Theorem~\ref{thm2.1}:}
We first observe that Eq.~(\ref{eq12}) follows from the above
definitions of $A_{\bf a}$ and $B_{\bf b}$.  Let $\Omega = [-3,
3n)^2$ and let $\kappa$ be the indicator function of the union of
the unit squares $\cup_{i=-2}^{3n} [i-1,i)^2$, lined up along the
main diagonal of $\Omega$, in symbols
\begin{eqnarray}
\kappa (u,v) & = & \sum_{i,j = - 2, -1, \ldots, 3n} \delta_{ij}
\cdot 1 \{i - 1 \leq u < i \} \cdot 1 \{j - 1 \leq v < j \}.
\label{eq15}
\end{eqnarray}
Here $1 \{\cdot\}$ denotes the indicator function of the set in
curly brackets and
\begin{eqnarray}
\delta_{jk} & = & \left\{
\begin{array}{ll}
1 & \mbox{ if } j = k \\
0 & \mbox{ if } j \neq k
\end{array} \right.%
\label{eq16}
\end{eqnarray}
denotes the Kronecker symbol.  On each of these $3n + 3$ unit
squares we place uniform mass, that comes from a product measure
on each of the squares, where the first factor only depends on the
setting $\textbf{a}$ and the second factor only depends on the
setting $\textbf{b}$.  Although this will make the mathematics
quite a bit more complicated, we can envision further experiments
where this feature of our construction may be of importance.  The
details are as follows. We define
\begin{eqnarray}
\sigma_{\bf a} (u) & = &
\begin{array}{ll}
|a_k| \cdot 1 \{-k \leq u < -k + 1\}  & k \in I_1 \\[3pt]
N_k (|a_1|) \cdot 1 \{k -1  \leq u < k\}  & k \in I_2 \\[3pt]
N_{k-n} (|a_2|) \cdot 1 \{k-1 \leq u < k\}  & k \in I_3 \\[3pt]
N_{k-2n} (|a_3|) \cdot 1 \{k-1 \leq u < k\}  & k \in I_4 \\[3pt]
0 \mbox{ elsewhere}
\end{array}%
\label{eq17}%
\end{eqnarray}
\begin{eqnarray} \tau_{\bf b} (v) & = & \begin{array}{ll}
|b_k| \cdot 1 \{-k \leq v < -k + 1\}  & k \in I_1 \\[3pt]
{1 \over 2} \psi_k (|b_1|) \cdot 1 \{k -1  \leq v < k\} & k \in
I_2 \\[3pt] {1 \over 2} \psi_{k-n} (|b_2|) \cdot 1 \{k-1 \leq v < k\} &
k \in I_3 \\[3pt] {1 \over 2} \psi_{k-2n} (|b_3|) \cdot 1 \{k-1 \leq u
< k\}  & k \in I_4 \\[3pt] 0 \mbox{ elsewhere}.
\end{array}
\label{eq18}%
\end{eqnarray}
The symbols $I_1, \ldots, I_4$ stand for $I_1 = +3, +2,+1$; $I_2 =
1, \ldots, n$; $I_3 = n+1, \ldots, 2n$; $I_4 = 2n +1, \ldots, 3n$.
We finally define the density $\rho_{\bf ab}^{(n)}$ by
\begin{eqnarray}
\rho_{\bf ab}^{(n)} (u,v) & = & \sigma_{\bf a} (u) \tau_{\bf
b} (v) \kappa (u,v)%
\label{eq19}
\end{eqnarray}
and the measure $\mu$ by having density $\rho_{\bf ab}^{(n)}$ with
respect to Lebesgue measure.

Hence we obtain from the above definitions the following integrals
needed for the calculation of the spin pair correlation function:
\begin{eqnarray}
\int_{[-3,0)^2} A_{\bf a} (u) B_{\bf b} (v) \rho_{\bf ab} (u,v) du
dv = \qquad \qquad \qquad \qquad \qquad & & \nonumber \\
- \sum_{k=1}^{3} |a_{k} \| b_{k}| {\mbox{sign}} (a_{k})
{\mbox{sign}} (b_{k}) \ = \ - {\bf a} \cdot {\bf b}. &&
\label{eq20}
\end{eqnarray}
Furthermore, the integral over the complement of the square
$[-3,0)^2$ vanishes, i.e.,
\begin{eqnarray}
\int_{\Omega \backslash [-3,0)^2} A_{\bf a} (u) B_{\bf b} (v)
\rho_{\bf ab} (u,v) du dv & = & 0
\label{eq21}%
\end{eqnarray}
which proves Eq.~(\ref{eq11}).

It remains to be shown that $\rho_{\bf ab}$ defines a measure
$\mu$ that is close to a probability measure, i.e., fulfills
Eq.~(\ref{eq10}).  For this, we consider the mass distribution
between the square $[-3, 0)^2$ and its complement. The amount of
mass $M_1$ distributed over $[-3, 0)^2$ is
\begin{eqnarray}
M_1 & = & \sum_{k=1}^3 | a_{k} \| b_{k} |.
\label{eq22}
\end{eqnarray}
The mass $M_2$ of $\Omega \backslash [-3,0)^2$ equals
\begin{eqnarray}
M_2 & = & {1 \over 2} \sum_{k=1}^3 \sum_{i=1}^n N_i (|a_{k}|)
\psi_i (|b_{k}|).%
\label{eq23}%
\end{eqnarray}
Thus the total mass distributed equals in view of Eq.~(\ref{eq14})
\begin{eqnarray}
M_1 + M_2 & = & \sum_{k=1}^3 | a_{k} \| b_{k} | + {1 \over 2}
\sum_{k=1}^3 \sum_{i=1}^n N_i (|a_{k}|) \psi_i (|b_{k}|)\nonumber \\
M_1 + M_2 & = & \sum_{k=1}^3 | a_{k} \| b_{k} | + {1 \over 2}
\sum_{k=1}^3 (|a_{k}| - |b_{k}|)^2 + \theta \cdot n^{-2} \nonumber \\
& & \qquad \qquad M_1 + M_2 \ \ = \ \ 1 + \theta \cdot n^{-2}
\label{eq24}
\end{eqnarray}
where $0 \leq \theta < 1/4$.

This completes the proof of the theorem which is the first
stepping stone of our construction of a suitable probability
measure.

Obviously, if instead of Eq.~(\ref{eq23}), we would define
\begin{eqnarray*}
M_2 & := & \sum_{K=1}^3 (|a_k| - |b_k|)^2,
\end{eqnarray*}
$\Omega = [-3,3]^2$ and would place the mass represented by these
three summands on any of the nine unit squares of $[0, 3)^2$, we
would produce a genuine probability measure, satisfying all
conclusions of Theorem~\ref{thm2.1}.

Finally, let $L \geq 1$, and $p_{\ell} \geq 0$ with $\sum_{\ell
=1}^L p_{\ell} = 1$.  For $0 \leq w < 1$, we define
\begin{eqnarray}
s (w) = (-1)^{\ell}, \quad & \frac{\ell-1}{L} \leq w <
\frac{\ell}{L} , \quad  & \ell = 1,
\ldots, L \nonumber \\
q (w) = p_{\ell} , \quad  & \frac{\ell-1}{L} \leq w <
\frac{\ell}{L}, \quad   & \ell = 1, \ldots, L.
\label{eq25}%
\end{eqnarray}
For $-3 \leq u$, $v < 3n$ and $0 \leq w < 1$ we define
$\widetilde{\Omega} = \Omega \times [0,1)$,
\begin{eqnarray*}
\widetilde{A}_{\bf a} (u,w) & = & A_{\bf a} (u) s (w) \\
\widetilde{B}_{\bf b} (v,w) & = & B_{\bf b} (v) s (w) \\
\widetilde{\rho}_{\bf ab} (u,v,w) & = & \rho_{\bf ab} (u,v) q (w).
\end{eqnarray*}
Then $\widetilde{A}_{\bf a}$ and $\widetilde{B}_{\bf b}$ only
depend on ${\bf a}$, $u$, $w$, and ${\bf b}$, $v$, $w$,
respectively. Moreover, they satisfy the properly modified
conclusion of Theorem~\ref{thm2.1}. This procedure extends
$\Omega$ to $\widetilde{\Omega}$ by adding as a factor the unit
interval $0 \leq w < 1$ with a given mass distribution.

\section{Definition of the Layers} \label{sec3}
We call the construction including the unit interval as factor,
given in Section~\ref{sec2} the first layer. To simplify the
notation we shall omit the $\sim$ sign from the $\Omega$, $A$,
$B$, and $\rho$.  As we noted in~\cite{hp} and \cite{hpp2} the
first layer does not yet provide a model that guarantees absence
of action at a distance. To achieve this goal we will now define a
system of layers.  These layers will be obtained by permuting all
the unit squares contained in $\Omega$, including the mass
distribution and the corresponding strips on which $A$ and $B$ are
defined.  In addition we shall duplicate the mass distribution of
each layer labeled $m$, labelling the duplicate layer $m'$.  On
the layer labeled $m$ the functions $A_{\bf a}^{(m)}$ and $B_{\bf
b}^{(m)}$ will remain unchanged. However, on the companion layer
labeled $m'$ we shall switch the signs of $A_{\bf a}$ and $B_{\bf
b}$, by setting $A_{\bf a}^{(m')} = - A_{\bf a}^{(m)}$ and $B_{\bf
b}^{(m')} = - B_{\bf b}^{(m)}$. As we observed in a recent
paper~\cite{apyref}, this simple modification of our original
construction encompasses all the desired features to achieve so
called parameter independence. We now present this program in
detail.

Think of each of the unit cubes $[i-1, i) \times [j-1) \times
[0,1)$, $i, j = -2, -1, \ldots, 3n$ together with their respective
mass distribution and the values of $A_{\bf a}$ and $B_{\bf b}$
defined on them as a unit ensemble.  We permute these unit
ensembles in the following way.  Choose three vertical strips
$[i-1, i) \times [-3, 3n) \times [0, 1)$, $i = -2, -1, \ldots, 3n$
and three horizontal strips $[-3, 3n) \times [j-1, j) \times
[0,1)$, $j = -2, -1, \ldots, 3n$.  These intersect in nine unit
cubes.  Place the three unit ensembles $[i-1,i) \times [i-1, i)
\times [0,1)$, $i = -2, -1, 0$ of the first layer onto three of
these nine unit cubes, such that each vertical and each horizontal
strip contains exactly one of these three unit ensembles of the
first layer, and move with them the vertical and horizontal strips
of the first layer.  This can be done in $36 \left( {3n +3 \atop
3} \right)^2$ different ways.  There are still $9n^2$ unit cubes
left to be assigned their unit ensembles. Choose $3n$ of them and
place on them these unit ensembles of the first layer where the
density was defined by $\frac{1}{2} N_i (|a_k|) \psi_i (|b_k|)$,
$i = 1, 2, \ldots, n$; $k = 1, 2, 3$.  This can be done in $\left(
{9n^2 \atop 3n} \right) (3n)$! different ways. Place the remaining
$9n^2 - 3n$ unit ensembles to fill up the empty spaces. They have
total mass 0.  This yields a grand total of
\begin{eqnarray}
36 \left( {3n + 3 \atop 3} \right)^2 \left( {9n^2 \atop 3n}
\right) (3n)! \label{eq26}
\end{eqnarray}
arrangements, which we call ``layers".  We call this number
$\frac{1}{2} N$.

At this point we exercise our option to let $p_{\ell} = p_{m
\ell}$ in Eq.~(\ref{eq25}), $m = 1, 2, \ldots, \frac{1}{2} N$
depend on the label of the layer.

In summary, on each layer the functions $A_{\bf a} (u, w; m)$ and
$B_{\bf b} (v, w; m)$ only depend on $(u,w)$ and $(v,w)$,
respectively. Each layer supports a measure $\mu_m = \mu_{{\bf
ab}m}^{(n)}$ satisfying
\begin{eqnarray*}
1 \leq \mu_m (\IR^3) < 1 + 1/n^2.
\end{eqnarray*}
Each measure $\mu_m$ has a density $\rho (u, v, w; m)$ with
respect to Lebesgue measure that can be written in the form
\begin{eqnarray*}
\rho (u,v,w; m) & = & \sigma_{\bf a} (u;m) \tau_{\bf b} (v;m)
\kappa (u, v; m) q (w;m)
\end{eqnarray*}
for $-3 \leq u$, $v < 3n$, $0 \leq w < 1$, with the obvious
interpretation of $\sigma_{\bf a}$, $\tau_{\bf b}$, $\kappa$ and
$q$. Moreover, by Eq.~(\ref{eq11}), we have for each $m = 1,
\ldots, \frac{1}{2} N$,
\begin{eqnarray}
\int_{\Omega^{(m)}} A_{\bf a} (u,w;m) B_{\bf b} (v,w;m) \rho
(u,v,w;m) du dv d w & = & - {\bf a} \cdot {\bf b}.%
\label{eq27}
\end{eqnarray}
As indicated at the beginning of this section, we shall duplicate
each layer so that at the end we will have a total of $N$ layers.
We renumber the original layers $m$ by the odd positive integers,
$2m-1$, say, $m = 1, 2, \ldots, \frac{1}{2} N$.  The companion
layer to the layer $2m-1$ will be assigned label $2m$, $m = 1, 2,
\ldots, \frac{1}{2} N$.  Each layer $2m-1$ and companion layer
$2m$ will be assigned density previously denoted $\rho (u, v, w;
m)$.  Each layer $2m -1$ will carry the functions, originally
denoted $A_{\bf a} (u,w;m)$ and $B_{\bf b} (v,w;m)$, whereas the
companion layer $2m$ will carry $-A_{\bf a} (u,w;m)$ and $-B_{\bf
b} (v,w;m)$, instead. Thus after renumbering the functions $A_{\bf
a}$ and $B_{\bf b}$ and the densities $\rho$ accordingly we have
for all $u$, $v$, $w$, and all $m = 1, 2, \ldots, \frac{1}{2} N$
\begin{eqnarray}
A_{\bf a} (u,w; 2m-1) + A_{\bf a} (u,w;2m) & = &  0 \nonumber \\
B_{\bf b} (\upsilon, w; 2m-1) + B_{\bf b} (v,w;2m) & = & 0
\label{new26}
\end{eqnarray}
and
\begin{eqnarray}
\rho (u,v,w; 2m-1) & = & \rho (u,v, w; 2m). \label{new27}
\end{eqnarray}
Moreover, Eq.~(\ref{eq27}) continues to hold for all $m = 1, 2,
\ldots, N$.  Of course, the equivalent effect had been achieved by
adding a fourth dimension $t$ and multiplying the original
functions $A_{\bf a}$ and $B_{\bf b}$ by a Rademacher function $r
(t)$. This was done in Section~5.3 of our paper~\cite{hp}.

With all the mathematical objects properly in place we now
finalize the second step of the construction of our model.  The
emission time of the $i$-th particle determines the measurement
time and thus the label $m$ where $m = 1, 2, \ldots, N$.  Recall
that the labelling variable $R$ has uniform distribution over the
integers $m = 1, 2, \ldots, N$, given by Eq.~(\ref{eq1}).

Apart from the random variable $R$ the construction so far is
plain calculus in $\IR^3$.  Only now we do define a realization of
the random variables $\Lambda_{{\bf a}t}^{*}$, $\Lambda_{{\bf
b}t}^{**}$, and $\Lambda_t$ by defining the conditional density of
$\Lambda_{{\bf a}t}^{*}$, $\Lambda_{{\bf b}t}^{**}$, and
$\Lambda_t$ given the random variable $R$ by
\begin{eqnarray}
\left.
\begin{array}{r} \mbox{Prob} (\Lambda_{{\bf a}t}^{*} \in [u, u
+ \Delta u), \Lambda_{{\bf b}t}^{**} \in [v, v+\Delta v),
\Lambda_t \in [w,w+\Delta w) \mid R = m)\,  \\[2pt]
= \rho (u, v, w;m)\Delta u \Delta v \Delta w  \quad \quad \quad \
\qquad \ \qquad \ \ \qquad
\\[2pt]
= \sigma_{\bf a} (u;m) \tau_{\bf b} (v;m) \kappa (u,v,m) q (w;m)
\Delta u \Delta v \Delta w \quad \\[2pt] - \infty < u, v < \infty, 0 \leq w <
1; m = 1, 2, \ldots, N.\qquad
\end{array} \right\}
\label{eq28}
\end{eqnarray}
This is the same as saying that the joint density of the four
random variables $\Lambda_{{\bf a}t}^{*}$, $\Lambda_{{\bf
b}t}^{**}$, $\Lambda_t$, and $R$ is given by
\begin{eqnarray}
\rho (u, v, w; m) \cdot \frac{1}{N}, \ \ && - \infty < u, v <
\infty, 0 \leq w < 1, m = 1, \ldots, N.
 \label{eq29}
\end{eqnarray}

A few remarks are in order. First, in previous write-ups we have
included mappings $f$ and $g$, to accommodate more general random
variables $\Lambda^*_{{\bf a}t}$ and $\Lambda_{{\bf b}t}^{**}$.
Obviously, this can be done here, too. Second, we changed the
model by defining $\rho_{\bf ab}$ to be the joint conditional
density of $\Lambda^*_{{\bf a}t}$ and $\Lambda^{**}_{{\bf b}t}$
given $R$, rather than by defining $\rho_{\bf ab}$ given by
Eq.~(\ref{eq19}) to be the joint conditional density of the mixed
parameters $\Lambda_{{\bf a}t}^1$ and $\Lambda_{{\bf b}t}^2$,
given $R$, as was done in \cite{hp,hpp2}. This makes for a more
streamlined presentation when the source parameter $\Lambda_t$ is
taken into account since obviously $\Lambda^1_{{\bf a}t}$ and
$\Lambda^2_{{\bf b}t}$, are functions of $\Lambda_t$ and the
station parameters $\Lambda^*_{{\bf a}t}$ and $\Lambda^{**}_{{\bf
b}t}$, respectively, and thus cannot be independent of $\Lambda_t$
(compare to condition (ii$^*$) in Section~\ref{sec1}). Hence, the
expression for the joint density corresponding to Eq.~(\ref{eq29})
would be more complicated.

We now discuss the stochastic dependence relations between the
four random variables $\Lambda^*_{{\bf a}t}$, $\Lambda_{{\bf
b}t}^{**}$, $\Lambda_t$, and $R$ that are direct consequences of
Eq.~(\ref{eq29}). First, the joint density of $\Lambda^*_{{\bf
a}t}$ and $\Lambda^{**}_{{\bf b}t}$ is given by
\begin{eqnarray*}
{1 \over N} \sum_{m=1}^N \sigma_{\bf a} (u; m) \tau_{\bf b} (v;m)
\kappa (u,v;m) \int_0^1 q (w;m) d w.
\end{eqnarray*}
Since the last integral equals 1, this reduces to, in view of
Eq.~(\ref{eq24}),
\begin{eqnarray*}
{1 \over (3n + 3)^2} \left( \sum_{k=1}^3 | a_k \| b_k | + {1 \over
2} \sum_{k=1}^3 \sum_{i=1}^3 N_i (|a_k|) \psi_i (|b_k|) \right) =
&& \\ {M_1 + M_2 \over (3n+3)^2} = {1 + \theta \cdot n^{-2} \over
(3n + 3)^2} &&
\end{eqnarray*}
with $0 \leq \theta < {1 \over 4}$. We conclude that the joint
density of $\Lambda^*_{{\bf a}t}$ and $\Lambda^{**}_{{\bf b}t}$ is
approximately uniform over the square $[-3, 3n)$ and, as a
consequence, equals the product of its two marginal densities
which are themselves approximately uniform over the interval $[-3,
3n)$.

We conclude that approximately:
\begin{description}
\item{(i)} $\Lambda^*_{{\bf a}t}$ and $\Lambda_{{\bf b}t}^{**}$
are independent random variables, and \item{(iv)} the
distributions of $\Lambda^*_{{\bf a}t}$ and $\Lambda^{**}_{{\bf
b}t}$ do not depend on ${\bf a}$, ${\bf b}$, and $t$.
\end{description}
Moreover, summation over $\ell = 1, 2, \ldots, L$ yields
\begin{eqnarray*}
\begin{array}{r}
\mbox{Prob} (\Lambda_{{\bf a}t}^{*} \in [u, u+ \Delta u),
(\Lambda_{{\bf b}t}^{**} \in [v,v+\Delta v) |R = m) \qquad \qquad \ \qquad \  \qquad \ \\
= \sigma_a (u;m) \tau_b (v;m) \kappa (u,v;m) \Delta u \Delta v,
\quad m = 1, \ldots, N, - \infty < u, v < \infty
\end{array}
\end{eqnarray*}
and integration over $- \infty < u$, $v < \infty$ yields
\begin{eqnarray*}
\mbox{Prob} (\Lambda_t \in [w, w + \Delta w) \mid R = m) \sim q
(w;m) \Delta w \quad 0 \leq w < 1, m = 1, \ldots, N.
\end{eqnarray*}
Thus by Eq.~(\ref{eq28}), we have approximately%
\begin{description}
\item{(ii)} Given $R$ the pair $(\Lambda_{{\bf a}t}^{*}$,
$\Lambda_{{\bf b}t}^{**})$ is conditionally independent of
$\Lambda_t$.
\end{description}
Also, approximately,
\begin{eqnarray*}
\begin{array}{r}
\mbox{Prob} (\Lambda_{{\bf a}t}^{*} \in [u, u + \Delta u),
(\Lambda_{{\bf b}t}^{**} \in [v, v + \Delta \sigma) \mid \Lambda_t
\in [w, w + \Delta w), R = m)  \\[5pt] = \displaystyle{\frac{\sigma_{\bf
a} (u;m) \tau_{\bf b} (\sigma; m) \kappa (u, v; m) \frac{1}{N} q
(w;m) \Delta u \Delta v \Delta w}{\mbox{Prob} (\Lambda_t \in [w, w
+ \Delta w) \mid R = m) P (R = m)}} \qquad \ \\[7pt]
= \sigma_{\bf a} (u;m) \tau_{\bf b} (v;m) \kappa (u,v;m) \Delta u
\Delta v \qquad \ \qquad \ \qquad \quad \quad
\\[5pt]
= \mbox{Prob} (\Lambda_{{\bf a}t}^{*} \in [u, u + \Delta u),
(\Lambda_{{\bf b}t}^{**} \in [v, v + \Delta v) \mid R = m).
\end{array}
\end{eqnarray*}
\begin{description}
\item{(ii$^*$)} Further if $p_{m \ell} = p_{\ell}$ independent of
$m$, i.e., if $\Lambda_t$ and $R$ are independent, then
$\Lambda_{{\bf a}t}^{*}$, $\Lambda_{{\bf b}t}^{**}$, and
$\Lambda_t$ are independent random variables.
\end{description}
Moreover, we obtain for the pair correlation integral
\begin{eqnarray*}
E \{A_{\bf a} B_{\bf b}\} & := & E \{A_{\bf a} (\Lambda_{{\bf
a}t}^{*},
\Lambda_t, R) B_{\bf b} (\Lambda_{{\bf b}t}^{**}, \Lambda_t, R) \} \\
& = & \frac{1}{N} \sum_{m=1}^N E \{ A_{\bf a}^{(m)} B_{\bf
b}^{(m)}\} = - {\bf a} \cdot {\bf b}
\end{eqnarray*}
by Eq.~(\ref{eq27}) and Eq.~(\ref{eq29}).  Since by construction
(see Eq.~(\ref{eq28}))
\begin{eqnarray*}
A_{\bf a}^{(2m-1)} + A_{\bf a}^{(2m)} = 0, \quad B_{\bf
b}^{(2m-1)} + B_{\bf b}^{(2m)} = 0, \quad m = 1, 2, \ldots,
\frac{N}{2},
\end{eqnarray*}
we obtain parameter independence first summing over $m$ to obtain
Eq.~(\ref{eq4}) and then by keeping the desired variables fixed
and by integrating over the remaining ones.

\section{A Model Based on the Poisson Process} \label{sec4}
The original experiment of Aspect et al.~\cite{eprex} took hours
to complete.  Currently, improvements of the technique have been
accomplished by various teams of experimenters~\cite{kwiat} and
the length of time it takes to perform these experiments has been
reduced substantially. The time between subsequent measurements is
still limited by the recovery (essentially a random process) of
the detectors between two measurements.

From the logistical angle the present section is designed to
replace the third paragraph of Section~\ref{sec1} and the parts of
Sections~\ref{sec2} and \ref{sec3} corresponding to it. Thus,
overall, the present section is a variant of that part of the
model dealing with generating the labels $m$. This will be done by
considering the waiting times between consecutive ``jumps" of a
Poisson process. Since we are entering more advanced mathematical
territory we present some of the relevant definitions and theorems
in a basic form rather than to send the reader searching through
the literature.

We first recall a few definitions from the theory of uniform
distribution $\mbox{mod} 1$.  For more details see~\cite{kuipers}
and \cite{drmota}. For a real number $x$, denote by $[x]$ the
integer part and by $\{x\} = x - [x]$ the fractional part of $x$.
Let $((x_i))_{i=1}^\infty$ be a sequence of real numbers.  For $k
\geq
1$ and $0 \leq \alpha < \beta \leq 1$ denote by%
\begin{eqnarray*}
A_{k} (\alpha, \beta) & := & \sum_{i \leq k} 1 (\alpha \leq
\{x_i\} < \beta)
\end{eqnarray*}
the number of elements $x_i$, $i \leq k$ such that their
fractional part $\{x_i\}$ is contained in a given interval
$[\alpha, \beta) \subset [0,1)$.  The sequence $((x_i))$ is called
uniformly distributed $\mbox{mod} 1$ if its discrepancy
\begin{eqnarray*}
D_{k} & := & \sup_{\alpha, \beta} \left| {1 \over k} A_{k}
(\alpha, \beta) - (\beta - \alpha) \right| \to 0,
\end{eqnarray*}
as $k \to \infty$, that is, if in the long run, each interval
$[\alpha, \beta)$ uniformly contains its proper share of points
$\{x_i\}$.  Equivalently, we could define $A_{k}$ by wrapping the
real axis around a circle of circumference 1 and count the number
of hits a given interval $[\alpha, \beta)$, now located on the
circle, receives from the sequence $((x_i))$, $i \leq k$, itself.

The standard mathematical model for spontaneous emissions of
particles, such as photons or electrons, is a Poisson process with
intensity $1 / \Theta$, say.  The waiting times $T_i$ between
successive emissions are independent identically distributed
random variables having exponential distribution with parameter
$\Theta$.  The following theorem is a special case of Theorem~2 of
H. Robbins~\cite{robbins}.

\begin{theorem}\label{thm2.2}
Let $((T_i))$ be the sequence of waiting times between consecutive
jumps of a Poisson process. Then with probability~1 the sequence
$((T_1 + \cdots + T_i))_{i=1}^\infty$ is uniformly distributed
$\mbox{mod} 1$.
\end{theorem}

\paragraph{Remark}  In fact, it follows easily from Robbins'
proof and the Erd\"{o}s-Tur\'{a}n inequality that with
probability~1 the discrepancy $D_{k}$ tends to zero at least with
speed $k^{- {1 \over 2}} (\log k)^3$.  There are more than a dozen
other papers extending Robbins' theorem.

In terms of weak convergence of probability measures
Theorem~\ref{thm2.2} can be reformulated in the following way.
(See e.g., Billingsley~\cite{bil95}, pp.~15--25.)  Let $\omega$ be
an element of the set $\Omega^*$ of probability 1 as in
Theorem~\ref{thm2.2}. Set $x_i = T_1 (\omega) + \cdots + T_i
(\omega)$. Let $P_{k}$ be the probability measure that assigns
point mass ${1 \over k}$ to each $\{x_i\}$, $1 \leq i \leq k$.  If
several $\{x_i\}$ coincide, let the mass add.  Then $P_{k}
\Rightarrow P$ in the sense of weak convergence.  Here $P$ denotes
Lebesgue measure on $[0,1)$.  For ease of presentation let us
reformulate Theorem~\ref{thm2.2} in terms of random variables. Let
$Y_{k}$ be a random variable defined on some probability space
such that
\begin{eqnarray*}
{\mbox{Prob}} (Y_{k} = \{x_i\}) = {1 \over k} && i = 1, 2, \ldots,
k
\end{eqnarray*}
and let $U$ be a random variable having uniform distribution on
$[0,1)$, i.e., ${\mbox{Prob}} (U \leq x) = x$, $0 \leq x \leq 1$.
Then Theorem~\ref{thm2.2} can be restated as follows.  For each
$\omega \in \Omega^*$, we have as $k \to \infty$
\begin{eqnarray*}
Y_{k} & \Rightarrow & U
\end{eqnarray*}
in the sense of weak convergence.  Let $N$ be defined in
Eq.~(\ref{eq26}).  For $m = 1, 2, \ldots, N$ define intervals
$I_{m}$, of length $1/N$ by
\begin{eqnarray*}
I_{m} & := & \left[ {m-1 \over N}, {m \over N} \right).
\end{eqnarray*}
Define a new random variable $R$ by setting
\begin{eqnarray*}
R=m & \mbox{    if     } & U \in I_{m}, m = 1, 2, \ldots, N.
\end{eqnarray*}
Then uniformly over all intervals $I_{m}$, $m = 1, 2, \ldots, N$,
we have
\begin{eqnarray}
{\mbox{Prob}} (R = m) & = & {\mbox{Prob}} (U \in I_m) = \lim_{k
\to \infty} {\mbox{Prob}} (Y_{k} \in I_{m}) = {1 \over N}.
\label{eq30}
\end{eqnarray}

As in Section~\ref{sec3}, suppose that the $i$-th pair of
particles has been emitted. Fix $\omega \in \Omega^*$. The time of
emission of the $i$-th pair equals $x_i = T_1 (w) + \cdots + T_i
(\omega)$. When reduced $\mbox{mod} 1$ the fractional part
$\{x_i\}$ determines a label $m$ with $1 \leq m \leq N$.  The
labelling variable $R$ has uniform distribution over the integers
$m = 1, 2, \ldots, N$ given by Eq.~(\ref{eq30}) or Eq~(\ref{eq1}).
However, at the time the pair of particles arrive at their
respective measuring stations, the devices may not yet be ready to
provide a measurement, because of recovery problems, etc. For $r =
1, 2$ we define the random variable $D_r$ by setting
\begin{eqnarray*}
\begin{array}{rcrl} D_r & = & +1 & \mbox{ if device is ready at
station } S_r \\ & = & 0 & \mbox{ if not}. \end{array}
\end{eqnarray*} Obviously, by stochastic independence,
\begin{eqnarray*} P (R=m|D_1, D_2) & = & P (R=m) = {1 \over N}
\qquad m = 1, 2, \ldots, N. \end{eqnarray*}

Hence, given that the devices at both stations are ready for
measurement, the labelling random variable $R$ still has uniform
distribution.

\section*{Acknowledgements}
We thank Salvador Barraza-Lopez for helpful discussions. The work
was supported by the Office of Naval Research N00014-98-1-0604.

\section*{Appendix 1}
The lemma is an immediate consequence of Theorem~\ref{thm2.2} of
Schumaker~\cite{sch81} for the special values of $m = 3$, $\ell =
0$, $r = n$, and the knots chosen to be $y_{\nu} = {\nu \over n}$
with $\nu = 0$, $\pm 1, \pm 2, \ldots$ Then by Schumaker's
equation (4.33) we have (dropping the fixed superscript~3 of
$N_i^3$):
\begin{eqnarray*}
(y-x)^2 & = & \sum_{i=-2}^n \phi_{i,3} (y) N_i (x) \mbox{ for all
} 0 \leq x \leq 1 \mbox{ and } y \in R.%
\end{eqnarray*}
Here
\begin{eqnarray*}
\phi_{i,3} (y) & =  (y - y_{i+1}) (y-y_{i+2})%
\end{eqnarray*}
and
\begin{eqnarray*}
0 \leq N_i (x) & \leq & 1 \mbox{ for all } x.%
\end{eqnarray*}
We now restrict $y$ to $0 \leq y \leq 1$.  Then for $-2 \leq i
\leq n$, we have $0 \leq \phi_{i,3} (y) < 2$ unless $y \in
[y_{i+1}, y_{i+2}]$.  Since we must avoid negative $\phi$, we set
$\phi = 0$ in this interval by defining new functions $\psi$:
\begin{eqnarray*}
\psi_i (y) & = & 0 \mbox{ if } y \in [y_{i+1}, y_{i+2}] \\ \psi_i
(y) & = & \phi_{i,3} (y) \mbox{ otherwise}.
\end{eqnarray*}
Since for $y \in [y_{i+1}, y_{i+2}]$ we have
\begin{eqnarray*}
| (y-y_{i+1}) (y-y_{i+2}) | & \leq & {1 \over 4n^2}%
\end{eqnarray*}
we have
\begin{eqnarray*}
0 & \leq & \sum_{i = -2}^n (\psi_i (y) N_i (x) - \phi_{i,3} (y)
N_i (x)) \ \  \leq \ \ {1 \over 4n^2}%
\end{eqnarray*}
because for any given $y$ and for all $x$, only one term in the
sum can be off by at most ${1 \over 4n^2}$. This proves the lemma.


\begin{thebibliography}{99}
\bibitem{foot1} In \cite{hp,hpp1,hpp2}, and \cite{hpp3}, we used lower
case letters for these random variables.
\end{thebibliography}
\end{document}